\documentclass[draft]{agujournal2019}
\usepackage{url} 
\usepackage{lineno}
\usepackage{soul}
\draftfalse

\journalname{Geophysical Research Letters}

\begin{document}
\title{BepiColombo at Mercury: Three flybys, three magnetospheres}

%
%

\authors{Hayley N. Williamson\affil{1}, Stas Barabash\affil{1}, Hans Nilsson\affil{1}, Martin Wieser\affil{1}, Jim M. Raines\affil{2}, Anna Milillo\affil{3}, Alessandro Aronica\affil{3}, Elisabetta DeAngelis\affil{3}, Yoshifumi Futaana\affil{1}, Daniel Heyner\affil{4}, Adrian Kazakov\affil{3}, Stefano Livi\affil{5}, Stefano Orsini\affil{3}, Kristin Pump\affil{4}, Daniel Schmid\affil{6}, Manabu Shimoyama\affil{1}, and Ali Varsani\affil{6}}

\affiliation{1}{Swedish Institute of Space Physics, Solar System Physics and Space Technology, Kiruna, Sweden}
\affiliation{2}{Department of Atmospheric, Oceanic and Space Sciences, University of Michigan, Ann Arbor, Michigan, USA}
\affiliation{3}{INAF/Istituto di Astrofisica e Planetologia Spaziale, Rome, Italy}
\affiliation{4}{Institut für Geophysik und Extraterrestrische Physik, Technische Universität Braunschweig, Braunschweig, Germany}
\affiliation{5}{Southwest Research Institute, San Antonio, TX, USA}
\affiliation{6}{Institut fuer Weltraumforschung, Graz, Austria}

\correspondingauthor{Hayley N. Williamson}{hayley.williamson@irf.se}

\begin{keypoints}
\item First observations of the equatorial magnetosphere from the SERENA-MIPA sensor
\item BepiColombo's first three Mercury flybys have different activity levels, resulting in different ion populations
\item Pitch angle analysis indicates the presence of precipitating bursty bulk flows in Mercury flybys 2 and 3
\end{keypoints}

%
\begin{abstract}
 We examine the first three BepiColombo Mercury flybys Using data from the Miniature Ion Precipitation Analyzer (MIPA), an ion mass analyzer in the Search for Exospheric Refilling and Natural Abundances (SERENA) package on the Mercury Planetary Orbiter (MPO) designed to study magnetospheric dynamics. These flybys all passed from dusk to dawn through the nightside equatorial region but were noticeably different from each other. In the first flyby, we observe a low latitude boundary layer and $\sim$1 keV ions near closest approach. For flybys 2 and 3 we see ions up to 14 keV in the same location, including freshly injected precipitating ions inside the loss cone. High time resolution data from flyby 3 show variations consistent with bursty bulk flows 10s long and occurring over $\sim30$s periods, the first such observation in this region. MIPA data demonstrate that high-energy injection processes are an important source of precipitation ions at Mercury.
\end{abstract}
\section*{Plain Language Summary}
Mercury has a magnetic field like Earth, but is much smaller and does not have an atmosphere, meaning its interaction with the solar wind can change quickly. BepiColombo is a joint ESA/JAXA mission on its way to Mercury and has made six flybys of Mercury. We use data from the Miniature Ion Precipitation Analyzer, a plasma instrument on BepiColombo, to look at Mercury's magnetosphere during the first three Mercury flybys. These three flybys all crossed roughly the same equatorial region of the magnetosphere on the nightside close to the planet, the first snapshots of this region. However, the MIPA data looks different for each flyby. Specifically, we see energetic ions in the second and third flybys that are not there in the first. By looking at the direction the ions are traveling and their variations in time, we conclude that these ions are part of a phenomenon called bursty bulk flow, which also happens at Earth, where the magnetic field farther away from the planet in the magnetotail drives plasma towards Mercury at high speeds. While we also see plasma going towards Mercury in the first flyby, it is at lower energies, indicating the magnetosphere was quieter then.


\section{Introduction}
Mercury, the smallest and least explored of the terrestrial planets, possesses an intrinsic dipole magnetic field like that of Earth, but on a much smaller scale. The magnitude of this intrinsic magnetic field is approximately 1\% of Earth's, creating a much more compact magnetosphere. Thus, while many of the same magnetospheric processes have been observed at Mercury as at Earth, those at Mercury are much faster, e.g. the Dungey circulation cycle occurs over a period of minutes, rather than hours \cite{Slavin2021}. Because Mercury additionally lacks a substantial atmosphere, magnetospheric processes such as solar wind precipitation are linked to the surface, where the particles can be backscattered or sputter heavier planetary species such as $\mathrm{Na}$ to create an exosphere. 

The Mercury Surface, Space ENvironment, GEochemistry, and Ranging (MESSENGER) mission revealed much about Mercury \cite{Solomon2018}, including the plasma environment as seen by the Fast Imaging Plasma Spectrometer (FIPS) \cite{Andrews2007}. One such revelation is the presence of dipolarization fronts, where newly reconnected magnetic flux travels planetward through the central plasma sheet into a more dipolar configuration. As occurs at Earth, these fronts are associated with bursty bulk flows, planetward-directed plasma flows with much higher speeds ($\sim 300$km/s) than typical convected plasma ($\sim 50$km/s) \cite{Dewey2018}, and energize protons to above 10 keV \cite{Sun2017}. These dipolarizations are roughly 10s long and frequently occur in a series of multiple fronts separated by 5--20s \cite{Dewey2020}, predominantly in the post-midnight region \cite{Sun2016}. As these fronts approach the planet, the dipolar region of Mercury's magnetic field is capable of braking and diverting the plasma away from the equatorial regions, shielding the surface from precipitation \cite{Glass2022}. However, this does not preclude bursty bulk flows from precipitating to the surface, as modeled solar wind precipitation maps show that the amount of low latitude precipitation is highly dependent on interplanetary magnetic field (IMF) direction \cite{Lavorenti2023a}: a southward IMF is more strongly coupled to Mercury's magnetic field, and hence the region of dipolar field lines will be smaller and equatorial regions are less protected at low altitudes.

MESSENGER's highly eccentric polar orbit allowed for low altitude observations in primarily the north polar region, limiting observations of plasma sheet features such as bursty bulk flows close to the planet. Thus the equatorial and southern latitudes at low altitudes were largely unobserved until the first Mercury flyby of the BepiColombo mission, a joint ESA-JAXA mission consisting of two spacecraft, the European Mercury Planetary Orbiter (MPO) and the Japanese Mio \cite{Benkhoff2021}. The spacecraft are currently in a `stacked' configuration on top of the Mercury Transfer Module until Mercury arrival at the end of 2026. While many instruments have their FOVs blocked in this configuration, the particle sensor suite Search for Exospheric Refilling and Emitted Natural Abundances (SERENA) has two ion sensors on MPO capable of making unhindered measurements during the flybys. Here, we focus on observations by the Miniature Ion Precipitation Analyser (MIPA), a small ion mass spectrometer designed to measure solar wind precipitation \cite{Milillo2020,Orsini2021}. During the first three Mercury flybys, MIPA was able to observe both steady-state convection and bursty bulk flow-like plasma at low altitudes, confirming the presence of energetic flows capable of precipitating to the surface. Additionally, the variation between flybys highlights the transient and dynamic nature of Mercury's magnetosphere. 

\section{Methods} \label{sec:methods}
\subsection{Instrument description}
MIPA, a member of a family of miniature, configurable plasma analyzers \cite{Wieser2016}, has an approximately 2$\pi$ hemispherical field of view and an energy range of 30 eV -- 14 keV. The number of pixels and energy bins are configurable onboard the instrument, resulting in varying resolutions, with a time of 20 s to scan all possible combinations with an electrostatic deflection system and cylindrical electrostatic analyzer. After passing through these systems, ions enter a small time-of-flight (TOF) cell and scatter on a start surface. The possible secondary electron emitted in the process is detected by a channel electron multiplier (START CEM). Particles then propagate with some probability to the stop surface, where they are detected by the STOP CEM. A coincidence event consists of a corresponding pair of start and stop signals. Given the known incident ion energy, the time between signals is used to find the particle velocity and, hence, mass per charge. The rate of START CEM detections is higher than that of coincidences events, so all data shown in this work are from the START CEM to achieve the highest sensitivity. For the first Mercury flyby (MFB1), MIPA was run in a 32 energy bin and 24 angular pixel configuration, while Mercury flyby 2 (MFB2) had 16 energy bins and 48 pixels. For Mercury flyby 3 (MFB3), the instrument observed a single angular direction and only scanned over energy, giving a limited field of view but much higher time resolution, only 375 ms for a full energy scan.

\subsection{Data processing}
MIPA is optimized for measurements of very high fluxes after orbit insertion and has a small geometrical factor \cite{Orsini2021}. As a result, MIPA measured low START count rates for the comparatively lower flyby fluxes. MIPA omnidirectional counts were therefore treated to remove noise and improve statistical certainty prior to conversion to differential flux. With a low count rate, the direction-integrated counts can be modeled as a Poisson distribution. In this distribution, the tail consists of unlikely events, single counts separated by large intervals in time. Thus, we assume the tail for each dataset contains only noise, and fit this portion of the Poisson distribution to obtain an average noise rate per flyby, which averages $3 \times 10^{-3} \ \mathrm{counts/s}$.

Counts are a discrete representation of the actual smoothly-varying particle fluxes. Particularly for low count rates, where a bin may have a single count over a given time, a direct conversion to differential flux, giving a time step with a high flux surrounded by zero flux, does not accurately represent the environment. Instead, we apply a smoothing process wherein single counts in an energy bin are spread in time over the half-interval to the prior and subsequent counts. This gives a most probable value for the actual number of particles during this time interval, which can then be converted to a statistically valid differential flux. The smoothing process is only performed on energy-time bins with a single count surrounded by empty bins; other bins are left unchanged. In effect, this creates data with a variable integration time depending on the number of counts, visible in Figs. \ref{fig:all3ET} and \ref{fig:mfb3_ca} as energy bins with the same flux value across multiple time steps. 

Finally, for noise subtraction, we use the average noise rate to determine the 68\% confidence interval for each time step and set the bin value to the center of the confidence interval, including for bins unchanged by the smoothing process. This process is only applied to the angularly-integrated data in Fig. \ref{fig:all3ET} and the single pixel data in Fig. \ref{fig:mfb3_ca}, as the full energy-angle-time distribution is not well-modeled by a Poisson distribution due to the widely varying sensitivity of the MIPA pixels.

We then convert counts to differential energy flux (Fig. \ref{fig:all3ET}, $\#/(cm^2 \, s \, sr \, eV/eV)$) and differential flux (Figs. \ref{fig:all3_erpa} and \ref{fig:mfb3_ca}, $\#/(cm^2 \, s \, sr \, eV)$). Differential energy flux is used in the energy-time spectrogram for ease of visualizing higher energies. Counts in each bin $C(E, \Omega)$ are converted to differential flux $j(E, \Omega)$ using
\begin{equation}
    j(E, \Omega) = \frac{C(E,\Omega)}{\tau \, E \, \eta_{sta} \, \mathcal{GF}(\Omega)}
\end{equation}
where $\tau$ is the integration time per bin, $E$ the center bin energy, $\eta_{sta}$ is the probability a particle is detected by the START detector, and $\mathcal{GF}(\Omega)$ is the energy-independent geometrical factor, for each pixel for the angularly-resolved data in Fig. \ref{fig:all3_erpa} and the average over the field of view for the smoothed energy-time data. Differential energy flux in Fig. \ref{fig:all3ET} is calculated with the same equation but omitting the factor $1/E$.

\subsection{Pitch angle and loss cone}
Pitch angle distributions are calculated using 1-minute averaged MPO-MAG data, which has been normalized to a unit vector. The MPO-MAG data is interpolated to the MIPA data time resolution and rotated from Mercury-Solar-Orbital (MSO) to MIPA instrument coordinates using SPICE \cite{acton1996pss}. Then for each time step, the angle between the boresight of each MIPA pixel and magnetic field direction is calculated:
\begin{equation}
    \alpha_{p} = \arccos{\frac{\vec{p} \cdot  \vec{B}_{mipa}}{|p| \, |B_{mipa}|}}
\end{equation}
where $\vec{p}$ is the MIPA pixel boresight vector and $\vec{B}_{mipa}$ is the magnetic field vector in MIPA coordinates. The pixels are then grouped into bins, for which a mean value is calculated. 

For a particle moving along a dipole field line the pitch angles satisfy the ratio
\begin{equation} \label{eq:pa}
    \frac{\sin^2{\alpha_1}}{\sin^2{\alpha_2}} = \frac{B_1}{B_2}
\end{equation}
where $\alpha_1,\ \alpha_2$ are the pitch angles at two altitudes along the field line and $B_1, \ B_2$ the magnetic field strength. The loss cone is the range of pitch angles $< \alpha_l$ with a mirror point, i.e. $\alpha = 90^{\circ}$ at the surface of Mercury. Using $\alpha_2 = 90^{\circ}$ in Eq. \ref{eq:pa} gives:
\begin{equation} \label{eq:loss_cone}
    \sin^2{\alpha_l} = \frac{B_{local}}{B_{surf}}
\end{equation}
We use the semi-empiric KTH22 model \cite{Pump2024} for the calculation of $B_{local}$ and $B_{surf}$ at the location of each time step by tracing the local field line to the surface, as the surface field must be obtained from a model. These values are then used in Eq.\ref{eq:loss_cone} to obtain the loss cone angle as a function of time.

\section{Results} 
\subsection{Flyby trajectories}
\begin{figure}
    \centering
    \includegraphics[width=0.5\textwidth]{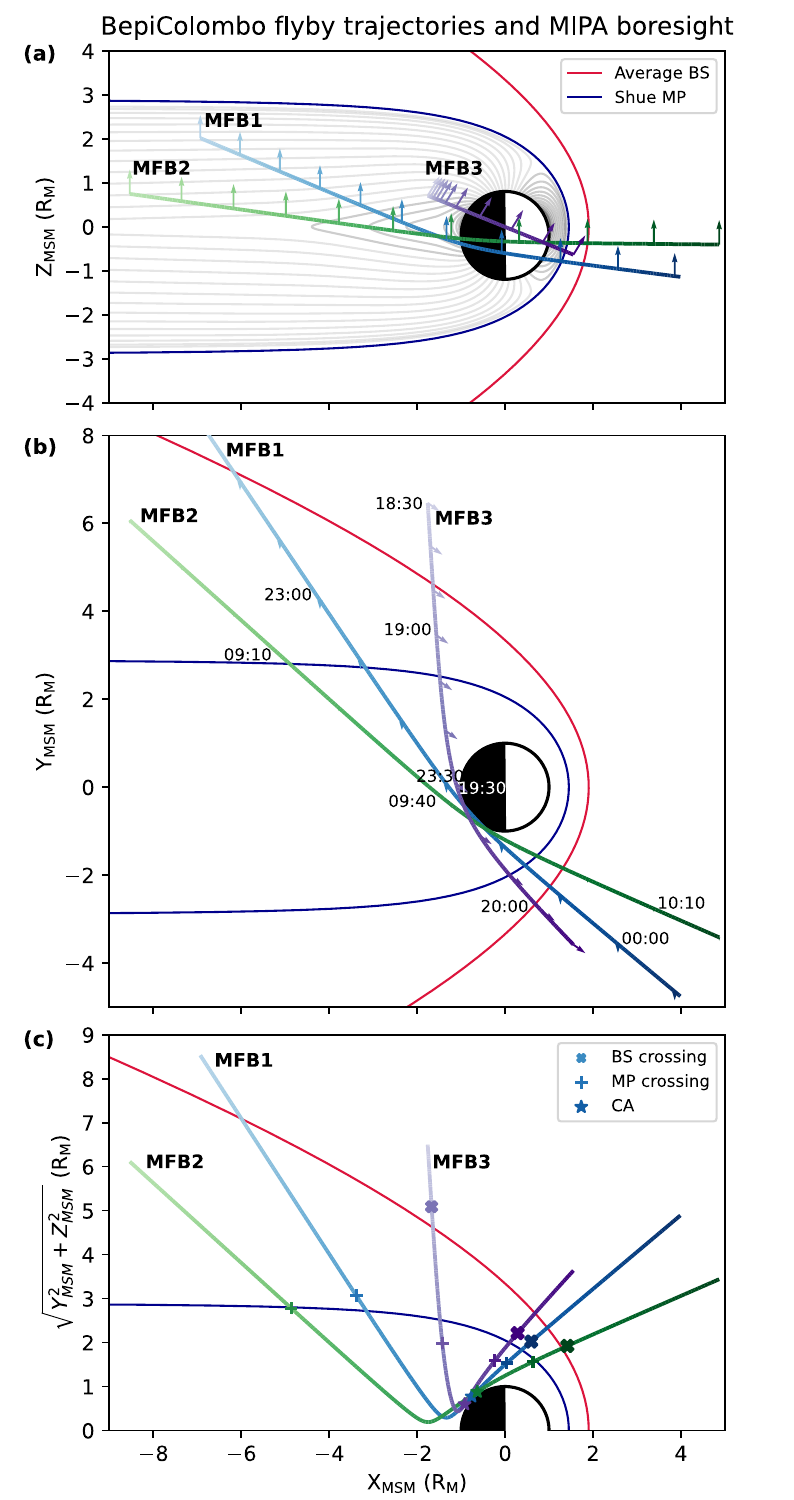}
    \caption{The BepiColombo trajectories for MFB1 (marked, blue), MFB2 (marked, green) and MFB3 (marked, purple) in MSM coordinates: (a) X-Z plane, (b) X-Y plane, and (c) X-$\rho$ plane. Each trajectory is shaded as a function of time and arrows indicate the MIPA boresight for each flyby, spaced 10 minutes apart. Panel (a) includes example field lines from the KT17 model.  Panel (c) includes the location of the magnetospheric boundaries observed by MIPA (bow shock = X, magnetopause = +) and closest approach (star). All panels include the Shue magnetopause model \cite<dark blue,>[]{shue1997jgr} and average bow shock (red) positions using parameters from \citeA{Winslow2013}.}
    \label{fig:trajectories}
\end{figure}
BepiColombo's first three Mercury flybys, on October 1, 2021 (MFB1); June 23, 2022 (MFB2); and June 19, 2023 (MFB3), had broadly similar trajectories, making comparisons of the observed ion populations possible. The flyby trajectories are shown in Mercury-Solar-magnetospheric (MSM) coordinates in Fig. \ref{fig:trajectories}, where X points towards the Sun, Z points towards the north pole, and Y completes the right-handed system, with the coordinate center offset by $Z = 0.2 \, R_M$ to coincide with the magnetic dipole. The panels show slices in the (a) X-Z plane, with fieldlines taken from the KTH17 model for display of far-tail field lines \cite{Korth_2017}; (b) X-Y plane, annotated with the time every 30 minutes; and (c) X-$\rho$ plane, where $\rho = \sqrt{Y^2 + Z^2}$. Panel (c) includes the positions of the bow shock and magnetopause crossings observed by MIPA (the inbound bow shock on MFB1 and MFB2 was crossed prior to MIPA switch-on). The arrows along the trajectories show the projection of the MIPA boresight onto the respective plane every 10 minutes, which for MFB1 and MFB2 is the center of the hemispherical field of view and for MFB3 the center of the single direction measured. The boresight was similar on all three flybys, primarily pointing north. While MFB3 is farther north than the first two, all three crossed through the nightside equatorial region from dusk to dawn, crossing the magnetic equator.

Comparison of the magnetopause and bow shock crossings for these three flybys with their average positions show a slight compression of the magnetosphere for MFB1 and MFB3. However, the boundaries are well within the typical distances seen by MESSENGER and Mariner 10 \cite{Winslow2013}, indicating the magnetosphere was not in a particularly extreme state during any of the flybys. Additionally, MFB3 occurred at lower heliocentric distances than the first two flybys, which is strongly correlated with compression of magnetospheric boundaries \cite<e.g.,>{zhong_boundaries_2015}. However, even minor changes in magnetospheric conditions are sufficient to change the observed ion populations, as we can see in the MIPA data.

\subsection{MIPA flyby overview}
\begin{figure}[ht]
    \centering
    \includegraphics[width=\linewidth]{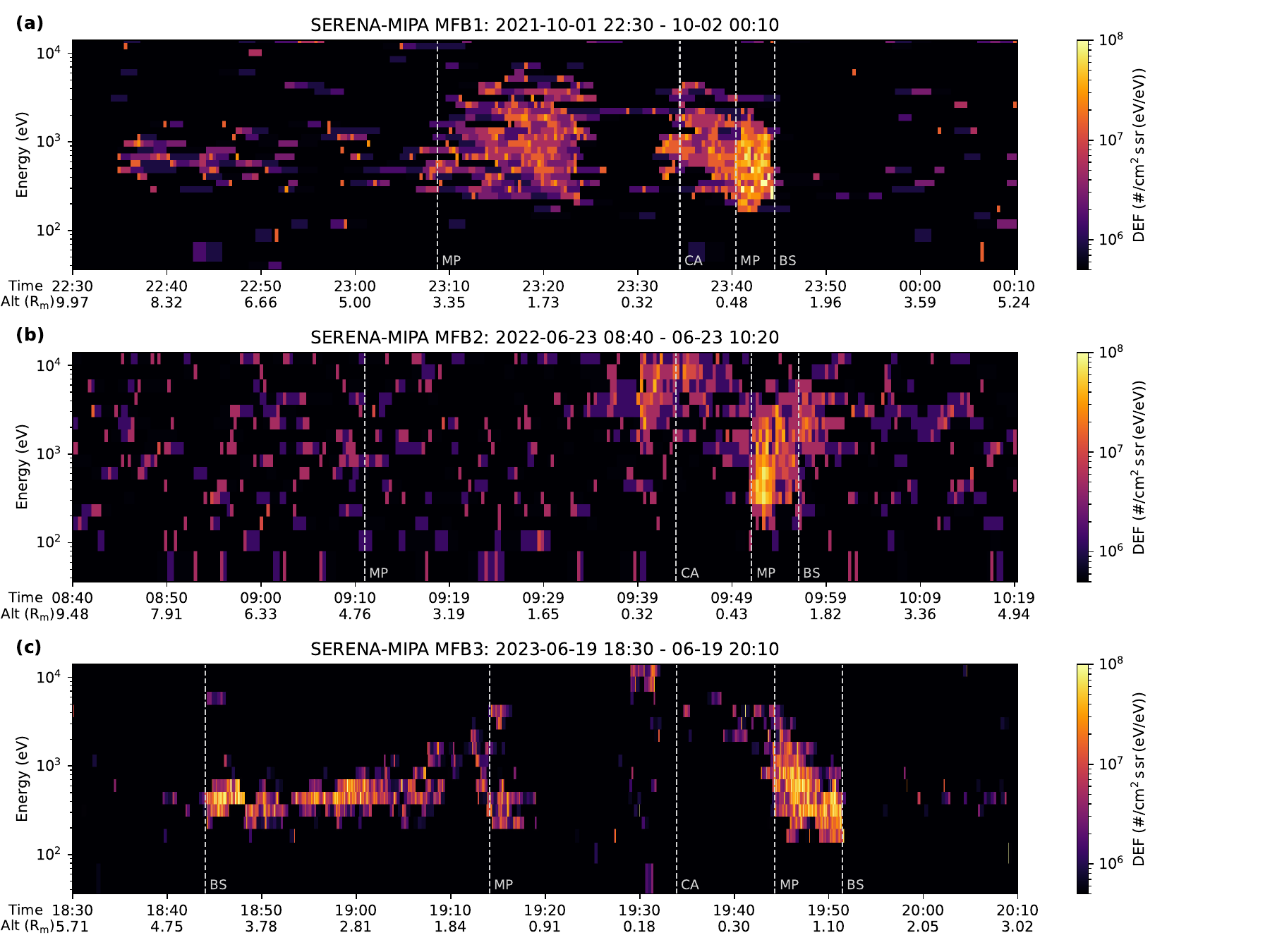}
    \caption{Energy-time spectrograms of the MIPA data for Mercury flyby 1 (a), 2 (b), and 3 (c). The x-axis is time in UTC + altitude in Mercury radii and the y-axis is energy measured by the instrument. The color indicates the omnidirectional differential energy flux in units of $\mathrm{\#/[cm^{2} \, s \, sr (eV/eV)]}$. The bow shock crossings, magnetopause crossings, and closest approach are marked in each panel as BS, MP, and CA respectively.}
    \label{fig:all3ET}
\end{figure}
Figure \ref{fig:all3ET} shows the MIPA differential energy flux as a function of energy and time. The time periods in the figure have been chosen so that closest approach for each flyby, marked with a dashed line and `CA', are aligned in the plot. We have also marked the magnetospheric boundary locations \cite{hewg_2024} with dashed lines, with the exception of the inbound bow shocks for MFB1 and 2. 

It is evident that, while the differential energy flux magnitude is similar across the flybys, the boundary locations and ion populations differ drastically. Beginning with the inbound (left) side of each plot, a weak, broad energy ion population consistent with a magnetosheath is apparent for MFB1. For MFB2 in Fig. \ref{fig:all3ET}(b), we cannot see a clear indication of an inbound magnetosheath, as magnetosheath plasma density was lower due to the MFB2 inbound trajectory crossing farther down the tail flank than MFB1 or MFB3 (Fig. \ref{fig:trajectories}(b)). In contrast, MFB3 in Fig. \ref{fig:all3ET}(c) passed through the near-terminator dusk flank, resulting in a particularly strong inbound magnetosheath signal. As MIPA was operated in a limited FOV mode for MFB3, a strong signal indicates a warm, broad ion population such as would be expected for a terminator magnetosheath. The energy is roughly similar to that of the MFB1 inbound magnetosheath, $\sim$ 500--1000 eV. 

The most significant difference between the flybys is inside the magnetopause, discussed in further detail below. The outbound magnetosheath, in the post-dawn region, is the most similar region between the flybys, with all showing decreasing energies when approaching the magnetopause, followed by high fluxes of warm, $\sim$300 eV -- 2 keV magnetosheath ions. However, we again see differences at the outbound bow shock crossing. For MFB1, we see a clear shock crossing, followed by little-to-no distinct signal. In MFB2, the shock crossing is much less distinct, consistent with a quasi-parallel bow shock \cite{hewg_2024}. The bow shock crossing is followed by an increasing energy dispersion to a population of a few keV ions moving both sunward and antisunward. Based on the flow directions and hybrid modeling of the flyby \cite{teubenbacher_solar_2024} we conclude this is the ion foreshock region.

After the outbound bow shock crossing of MFB3, which is clearly delineated similarly to MFB1, there is a sporadic signal at around 500 eV. Comparisons with the SERENA Planetary Ion CAMera (PICAM) sensor, which has routinely observed the solar wind throughout the cruise phase \cite{Alberti2023,Rojo2025}, indicate this is the solar wind. The sunward direction is not within the MIPA FOV for this particular spacecraft attitude and selected pixels. However, an aberration of 10$^{\circ}$ for the relatively low velocity solar wind ($\sim$270 km/s), combined with a non-zero angular response outside the pixel full-width-half-maximum, allow MIPA to observe this low flux $\sim380$ eV population. 

\subsection{Inside the magnetosphere}
Three distinct populations can be observed when comparing the spectrograms in Fig. \ref{fig:all3ET} inside the magnetopause. In Fig. \ref{fig:all3ET}(a) from $\sim$23:10--23:25 is the broad low latitude boundary layer (LLBL) of MFB1, a region of warm, high-density magnetosheath-like plasma just inside the equatorial magnetopause, followed by entrance into the central plasma sheet. The MFB1 LLBL was previously described in \cite{Orsini2022}. MFB2 lacks an LLBL, while MFB3 occurred at higher latitudes, although the energy-dispersed population inside the magnetopause has been theorized to be an LLBL \cite{Hadid2024}.

\begin{figure}
    \centering
    \includegraphics[width=\textwidth]{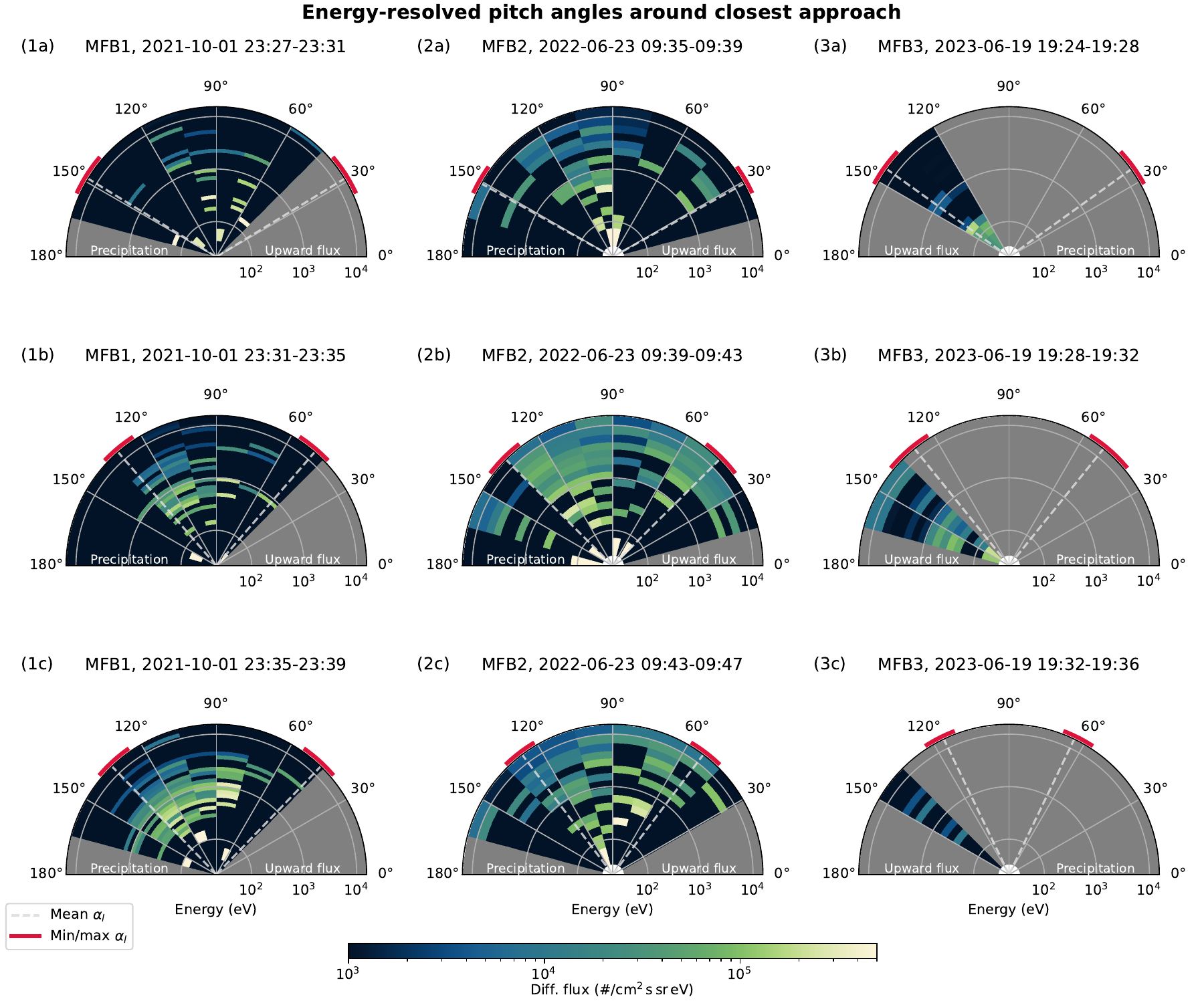}
    \caption{The energy-resolved pitch angle distributions in differential flux for all three flybys for a total of twelve minutes around closest approach. The first row are for four minutes prior to CA, the second row for four minutes centered on CA, and the third row for the subsequent four minutes before the magnetopause crossing. The first column (1a-c) are the distributions for MFB1, second column (2a-c) MFB2, and third column (3a-c) MFB3. Gray indicates pitch angles outside of the MIPA FOV. The mean loss cone boundary is marked with a white dashed line and the minimum/maximum loss cone boundary angles for each four minute period are indicated by the red line at the axis. The precipitating and upward directions are labeled in each plot.}
    \label{fig:all3_erpa}
\end{figure}

Near closest approach, we observe significant differences in ion energies between flybys. MFB1 shows $\sim 1$ keV plasma, while MFB2 (b) and MFB3 (c) passed through a high energy ion population up to the maximum MIPA energy of 15 keV. A sparse $<1$ keV population is also visible in MFB3, likely cold ions mostly outside of the FOV. To analyze this region further, we calculate the energy-resolved pitch angle distributions as described in Section \ref{sec:methods}. We take the mean of the distributions in three 4-minute periods: before, at, and after closest approach in each flyby, for better counting statistics, shown in Fig. \ref{fig:all3_erpa}.

We include the loss cone boundary ($\alpha_l$) calculated from the KTH22 model as white dashed lines in each panel of Fig. \ref{fig:all3_erpa}. Particles with pitch angles outside $\alpha_l$ will mirror; particles with pitch angles $0^{\circ}<\alpha<\alpha_l$ in the northern hemisphere (MFB3) and $180^{\circ}-\alpha_l<\alpha<180^{\circ}$ in the south (MFB1 and 2) will precipitate, while the other side of the loss cone contains upwelling ions. As Mercury's magnetosphere is small and the flybys are at low altitudes, $\alpha_l$ is significantly larger than at Earth. Because the spacecraft traveled a considerable distance over a 4-minute period, the range of $\alpha_l$ for each period is marked in red on the outer axis. 

The first two periods for MFB1, Fig. \ref{fig:all3_erpa}(1a-b), do not contain any significant fluxes inside the precipitating loss cone, only bouncing ions. However, Fig. \ref{fig:all3_erpa}(1c), from 23:35--23:39, shows precipitating energy-dispersed ions, with energy decreasing as pitch angle increases due to higher energy ions (i.e. shorter bounce time) precipitating more quickly than those at lower energies. This is consistent with substorm-injected electrons observed during the same time period by the Mio MPPE instrument \cite{Aizawa2023}. As MFB1 was in the southern hemisphere, fluxes are higher for $\alpha > 90^{\circ}$ (towards the planet).

MFB2 observations (second column) show consistently wide pitch angle distributions, although fluxes increase at closest approach in Fig. \ref{fig:all3_erpa}(2b). Indeed, the distribution is broader at energies above 1 keV and includes a significant component inside the loss cones for the entire 12 minutes. Ions $<1$ keV have a pitch angle distribution close to $90^{\circ}$ and fluxes comparable to Fig. \ref{fig:all3_erpa}(1b-c) from MFB1; thus, this is likely the same quasi-stable bouncing population. However, the ions $>1$ keV must be freshly injected via an energetic process that does not occur during MFB1. The ions in MFB3 are likewise inside the loss cone for a period of at least 8 minutes, Fig. \ref{fig:all3_erpa}(3b-c), with characteristics of both MFB2 (energy $>1$ keV) and MFB1 (Fig. \ref{fig:all3_erpa}(1c) $< 1$ keV) precipitation.

\begin{figure}
    \centering
    \includegraphics[width=\textwidth]{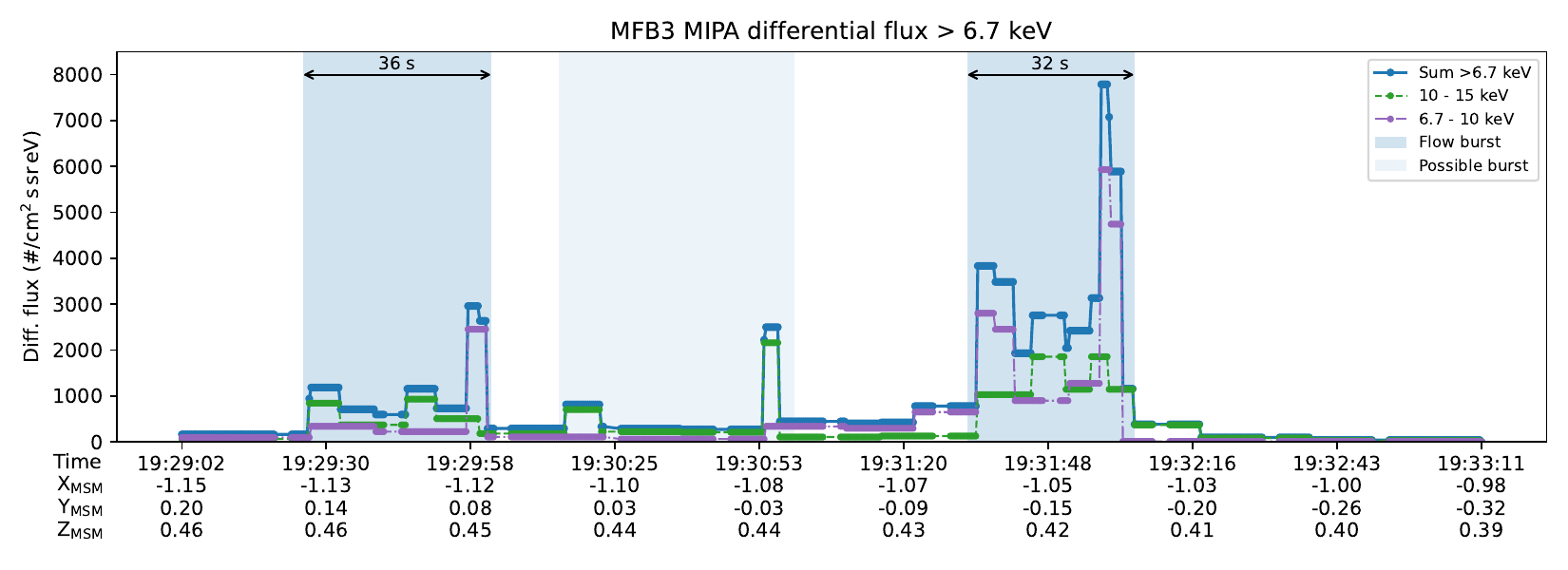}
    \caption{A zoom-in of the MIPA data for the MFB3 closest approach for the top two MIPA energy bins, 6.7--15 keV. The x axis shows time and MSM coordinates and the y axis differential flux ($\#/(cm^2 \, s \, sr \, eV)$). The solid blue line and markers indicate total differential flux for energies above 6.7 keV, and the individual energy bins are indicated with the green dashed line (10--15 keV) and purple dash-dotted line (6.7--10 keV). Shaded regions indicate flow bursts annotated with the duration, with a possible burst in a lighter shade.}
    \label{fig:mfb3_ca}
\end{figure}

While the data from MFB1 and 2 have the MIPA nominal 20 s time resolution, the fixed-deflection mode of MFB3 takes only 375 ms per energy scan. Fig. \ref{fig:mfb3_ca} shows the high time resolution differential flux around closest approach in more detail for energies  $\geq 6.7$ keV. The data have been smoothed as described in Section \ref{sec:methods}, decreasing the time resolution somewhat. However, short variations on the order of a few seconds in the high energy fluxes are still observed, and longer bursts of around 30 s are clearly visible (shaded regions). As discussed below, this variation is consistent with the timing of dipolarization fronts.

\section{Discussion}
The pitch angle distributions in Fig. \ref{fig:all3_erpa} show that the MFB2 and 3 represent a significantly more active magnetosphere than that observed in MFB1. While plasma circulation is continuous as part of the Dungey cycle, the dynamic nature of Mercury's magnetosphere means the nature of the reconnection-driven structures and plasma sheet can vary drastically. The $\sim$1 keV ions in MFB1 are consistent with a `quiet' plasma sheet \cite{Dewey2018,Sun2017}, but the $\geq$10 keV ions in MFB2 and 3 require an active magnetospheric state, with recurring dipolarization fronts and associated bursty bulk flows. The lower activity in MFB1 is exemplified by the presence of a LLBL, which is correlated with slower reconnection rates and hence a quieter magnetosphere \cite{Liljeblad2015}, while the substorm injection in panel (1c) of Fig. \ref{fig:all3_erpa} at typical plasma sheet energies is indicative of steady reconnection-driven convection.

While multi-keV ions in MFB2 with $\sim 90^{\circ}$ pitch angles could potentially be part of a ring current \cite{Shi2022,Zhao2022,Hadid2024}, this cannot fully explain the MIPA data. Ions inside the loss cones require a continuous, fresh injection of energetic particles that persists for up to 12 minutes in MFB2 and more than four minutes in MFB3. The presence of these freshly injected ions in the post-midnight region and energies of $>10$keV match MESSENGER observations of dipolarizarion fronts and bursty bulk flows at higher altitudes \cite{Sun2016,Dewey2020,Sun2017}. Additionally, the series of multiple $\sim30$s bursts seen in Fig. \ref{fig:mfb3_ca} fits a typical dipolarizarion front pattern \cite{Dewey2020}. The unavailability of full-resolution magnetometer data makes it difficult to confirm the presence of a cluster of dipolarization fronts driving bursty bulk flow. The consistency with MESSENGER observations lead us to infer this as the cause. The presence of both steady convection and higher energy reconnection structures were observed under extreme solar wind conditions in previous studies \cite{Sun2020a}; these flybys show that they likely are common features of the magnetosphere.
\begin{table}[]
\caption{A summary of the upstream parameters (solar wind velocity and IMF direction) and conditions observed (low latitude boundary layer, bursty bulk flow, and magnetospheric state) for the three flybys. Solar wind velocity is taken from PICAM data and IMF $\mathrm{B_Z}$ from Mio-MGF and MPO-Mag data in the cited papers.}
\label{tab:summary}
\begin{tabular}{lccccc}
\textbf{Flyby} & \textbf{$\mathrm{v_{sw}}$ (km/s)} & \textbf{$\mathrm{B_z}$} &  \textbf{LLBL} & \textbf{BBF} & \textbf{MS state} \\ \hline
MFB1$^{a}$ & $\sim 300$ & $> 0$  & Yes & No & Quiet \\ 
MFB2$^{b}$ & 450 & $< 0$  & No & Yes & Active \\ 
MFB3$^{b,c}$ & 330 & $< 0$  & N/A & Yes & Active \\ \hline
\multicolumn{2}{l}{$^{a}$\citeA{Alberti2023}} &
\multicolumn{2}{l}{$^{b}$\citeA{Teubenbacher2025a}} &
\multicolumn{2}{l}{$^{c}$\citeA{Rojo2025}}
\end{tabular}
\end{table}

To understand the driver of the change in activity level, we compare the upstream conditions from PICAM and the two BepiColombo magnetometers and the magnetospheric observations made by MIPA in Table \ref{tab:summary}. The last column, magnetospheric state, is based on the observed features described above, such as presence of a LLBL and bursty bulk flow. A notable difference is that, in addition to a lower $v_{sw}$, the IMF was northward for MFB1. The reconnection rate at Mercury is not strongly dependent on IMF direction \cite{DiBraccio2013} and previous studies of bursty bulk flows do not explore an association with IMF $B_z$ \cite<e.g.,>{Dewey2018,Dewey2020,Sundberg2012}. However, the ability of Mercury's magnetic field to shield the near-equatorial regions from precipitation is likely related to the IMF $B_z$ direction, as southward IMF creates a more open magnetic field configuration and smaller nightside dipolar region \cite{Lavorenti2023a}. Hence, the northward IMF during MFB1 likely resulted in a larger dipolar region and thicker plasma sheet that did not generate bursty bulk flows, but maintained substorm injections at plasma sheet energies. On the contrary, MFB2 and 3 had a more open magnetic field configuration that could both generate the dipolarization-front driven bursty bulk flows and compress the nightside dipolar region to lower latitudes, leading to the equatorial high energy precipitating ions we observe. Future orbital phase data from MIPA will help to reveal the frequency with which each type of injection occurs, as well as the connection to upstream conditions measured by Mio. 

\section{Conclusion}
We compare MIPA data for the first three BepiColombo Mercury flybys. There is significant variation in the plasma populations seen, despite all three passing through the nightside equatorial region. Comparison of the ions measured near closest approach suggest that MFB1 occured in a quiet magnetosphere with a low latitude boundary layer. However, MFB2 and MFB3 show ongoing, high energy plasma injection persisting over several minutes. Data from MFB3 show time variations of the $>6$keV ion fluxes on the order of 30s, meaning the high energy ions seen are consistent with a bursty bulk flow and a more active magnetosphere. Together, the data suggest the presence of a series of dipolarization fronts in MFB2 and 3 driving bursty bulk flow while the nightside dipolar region was compressed by a southward IMF, leading to high energy ion precipitation.
%
%
\section*{Open Research Section}
All MIPA data used to produce the figures in the paper, as well as the software used to process and plot the data, are currently available at \url{https://data.irf.se/williamson2025grl} or \url{https://gitlab.irf.se/hayley/mipa_flybys123} for the purposes of peer review. The MAG data used to calculate pitch angle is currently proprietary, so the authors have provided the already-processed pitch angle distributions, as well as the functions used to calculate them. The KTH22 model is available at \url{https://github.com/Krissy37/KTH-V10.git}.

\acknowledgments
SERENA is supported by the Italian Space Agency (ASI) and by the Italian National Institute of Astrophysics (INAF) under SERENA agreement no. 024-66-HH.0 “Attività scientifiche per il Payload SERENA su Bepi-Colombo, relative alla fine della fase di crociera e fase operativa”. The MIPA sensor was developed by the Swedish Institute of Space Physics with support by the Swedish National Space Agency and the University of Bern. We thank all of the SERENA team for their support and the developers of the KT17 and KTH22 models for the use of their tools. We also thank D. Heyner for use of the MPO-MAG data. H. Williamson and other Swedish Institute of Space Physics authors are supported by Swedish National Space Agency grant 2018-00067 for long-term support of the MIPA sensor. The authors declare there are no conflicts of interest for this manuscript.


%
%



\bibliography{Mercury}

%
%
%
%
%

\end{document}